\documentclass[twocolumn,superscriptaddress,amsfonts,amssymb,amsmath]{revtex4}

\usepackage{graphicx}%
\usepackage{dcolumn}%
\usepackage{amsmath}%

\begin{document}

\date{Oct 5th, 2001}

\title{Synthesis, Crystal Structure and Magnetic Properties of the 
Linear-Chain Cobalt Oxide Sr$_5$Pb$_3$CoO$_{12}$}

\author{K. Yamaura}
\email[E-mail:]{YAMAURA.Kazunari@nims.go.jp}
\homepage[Fax.:]{+81-298-58-5650}
\affiliation{Advanced Materials Laboratory, National Institute for Materials
Science, 1-1 Namiki, Tsukuba, Ibaraki 305-0044, Japan}
\affiliation{Japan Science and Technology Corporation, Kawaguchi, Saitama 
332-0012, Japan}

\author{Q. Huang}
\affiliation{NIST Center for Neutron Research, National Institute of Standards
and Technology, Gaithersburg, Maryland 20899}
\affiliation{Department of Materials and Nuclear Engineering, University of 
Maryland, College Park, Maryland 20742}

\author{E. Takayama-Muromachi}
\affiliation{Advanced Materials Laboratory, National Institute for Materials
Science, 1-1 Namiki, Tsukuba, Ibaraki 305-0044, Japan}

\begin{abstract}
The novel spin-chain cobalt oxide Sr$_5$Pb$_3$CoO$_{12}$ [{\it P}\={6}2{\it 
m}, {\it a} = 10.1093(2) \AA~and {\it c} = 3.562 51(9) \AA\ at 295 K] is 
reported.
Polycrystalline sample of the compound was studied by neutron diffraction (at
6 and 295 K) and magnetic susceptibility measurements (5 to 390 K). 
The cobalt oxide was found to be analogous to the copper oxide 
Sr$_5$Pb$_3$CuO$_{12}$, which is comprised of magnetic-linear chains at 
inter-chain distance of 10 \AA.
Although the cobalt oxide chains ($\mu_{\rm eff}$ of 3.64 $\mu_{\rm B}$ per 
Co) are substantially antiferromagnetic ($\theta_{\rm W} =$ -38.8 K), neither
low-dimensional magnetism nor long-range ordering has been found; a 
local-structure disorder in the chains might impact on the magnetism.
This compound is highly electrically insulating.

\end{abstract}

\maketitle

\pagebreak

\section{Introduction}

In order to reveal nature of correlated electrons in condensed matters and to
construct advanced models for those, one dimensional (1D) electronic system 
as a basis of the models has been subjected for both experimental and theoretical 
investigations.
In the experimental part, a variety of quantum phenomena has been indeed found
for the quasi-1D compounds in past few decades; Tomonaga-Luttinger-type 
electric conductivity for BaVS$_3$ \cite{PRB94MN}, Peierls instability for 
CuGeO$_3$ \cite{PRL93MH}, charge- and spin-density waves for (TMTSF)$_2$AsF$_6$
and TTF-TCNQ \cite{PRB82KM,JPSJ76SK}, spinon-holon separation for SrCuO$_2$ 
\cite{PRL96CK}, Haldane gap for Ni(C$_2$H$_8$N$_2$)$_2$NO$_2$(ClO$_4$) \cite{PRL89AY}, 
and superconductivity for (TMTSF)$_2$PF$_6$ \cite{JPL80DJ}. 
These characteristic phenomena might reflect the nature of the correlated 
electrons, and intensive research on the materials played an important 
role to steadily advance understanding of the nature.
We have recently been exploring novel quasi-1D compounds in order to find 
additional systems showing correlations among their magnetic, electronic 
transport properties, and crystal structure.

The linear-chain cobalt oxide Sr$_5$Pb$_3$CoO$_{12}$ was recently discovered
in a course of studies of quasi-1D magnetic materials and 
electrical-carrier-doped those \cite{PRB01KY,SSC00KY,JSSC99KY,JSSC90JSK,JSSC91TGNB}.
The polycrystalline sample of the compound was obtained by high-temperature 
solid-state reaction, and then subsequently investigated by x-ray diffraction,
magnetic susceptibility, and thermogravimetric analysis (TGA) studies.
Crystal structure was investigated in detail by powder-neutron-diffraction 
at 6 and 295 K.
The compound was found isostructural to the 1D antiferromagnetic copper oxide 
Sr$_5$Pb$_3$CuO$_{12}$ [{\it P}\={6}2{\it m}, $a =$ 10.1089(6) \AA~and $c =$
3.5585(2) \AA], in which distorted CuO$_4$ units are linearly connected by 
sharing those conner-oxygen \cite {PRB01KY,JSSC90JSK,JSSC91TGNB}.
Spin-singlet ground state, which potentially results from linear-alternation
of quantum spins, was indeed suggested at the composition Sr$_5$Pb$_{2.6}$Bi$_
{0.4}$CuO$_{12}$ \cite{PRB01KY}. 
Although the title antiferromagnetic compound Sr$_5$Pb$_3$CoO$_{12}$ was 
therefore expected to present one dimensionally anisotropic magnetism, however
Curie-Weiss(CW)-type magnetism was eventually observed even at low temperature 
($>$ 5 K).
The rather isotropic magnetic property probably indicates a lack of 
1D-magnetic uniformity, which may be affected by high degree of structural disorder in 
chains, as in the Bi-nondoped copper analogue \cite{PRB01KY}.
In this paper, the crystal structure and magnetic properties of the novel 
chain compound are reported and compared with those of the analogous 
copper oxides.

\section{Experimental}

The samples of the cobalt oxide were prepared as follows. 
A mixture of pure ($\ge$ 99.9 \%) and fine powders of SrCO$_3$, PbO, and 
Co$_3$O$_4$ (Sr:Pb:Co = 5:3:1 molar ratio, $\sim$7 g) was placed in a dense-alumina crucible
with a cap and heated in air at 750 $^\circ$C for 19 hours.
The sample was quenched at room temperature and then ground. 
Thereafter, the powder was reheated at 850, 900, and 950 $^\circ$C in turn for
66 hours in total.
The preheated powder was then molded into several pellets, followed by heating
at 970 $^\circ$C for 73 hours in air.
Subsequently, some of the sintered pellets were annealed at 500 $^\circ$C in mixed gas,
20 \% oxygen in argon, at 100 MPa for 5 hours in a commercial apparatus 
(hot-isostatic-pressing system, developed by KOBE STEEL, LTD).
To qualitatively analyze the polycrystalline samples, x-ray diffraction at
room temperature with CuK$_\alpha$ radiation was employed.
The x-ray apparatus (RINT-2000 system, developed by RIGAKU, CO) was equipped
with a graphite monochromator on the counter side and an auto-divergence-slit
system.
The oxygen content of the samples was measured in a commercial TGA apparatus 
(PYRIS 1, developed by PerkinElmer, Inc) by heating small amount of each 
sample (powder, $\sim$20 mg) in 3 \% hydrogen in argon at a heating rate of 5
$^\circ$C per minute to 700 $^\circ$C and holding for 6 hours.

To investigate the crystal structure further, the sample annealed in the 
compressed gas was again studied by neutron diffraction.
The neutron data at 6 and 295 K were obtained by the BT-1 high-resolution 
powder diffractometer at the NIST Center for Neutron Research.
A Cu(311) monochromator was employed to produce a coherent neutron beam
($\lambda = 1.5401$ \AA) with 15$^{\prime}$, 20$^ {\prime}$, and 7$^{\prime}
$ collimators before and after the monochromator, and after the sample, 
respectively. 
The neutron diffraction patterns were measured between 8 and 160 
degrees in diffraction angle at 0.05 degrees step.
With the neutron profiles, crystal structure parameters were refined to a high
degree of agreement by Rietveld calculations on the program GSAS \cite 
{LANLR90ACL}\@.
The total numbers of reflection and data point were 172 and 3039, respectively.
Neutron scattering amplitude in the calculations were set 0.702, 0.940, 0.253,
and 0.581 ($\times$10$^{-12}$) cm for Sr, Pb, Co, and O\@, respectively \cite
{LANLR90ACL}.

The magnetic properties of the samples were studied by a commercial apparatus
(MPMS system) between 5 and 390 K, developed by Quantum Design, Inc. 
The magnetic susceptibility data were collected at 50 kOe on cooling, and 
magnetization curves were recorded between -55 and 55 kOe after cooling the 
each sample at 5 and 150 K. 
All the pellets thus obtained were too electrically insulating to be subjected
for further electronic transport measurements; beyond the 10 M$\Omega$ limit
of a conversional two-terminal tester at room temperature.

\section{Results and Discussions}

\begin{figure*}
\includegraphics{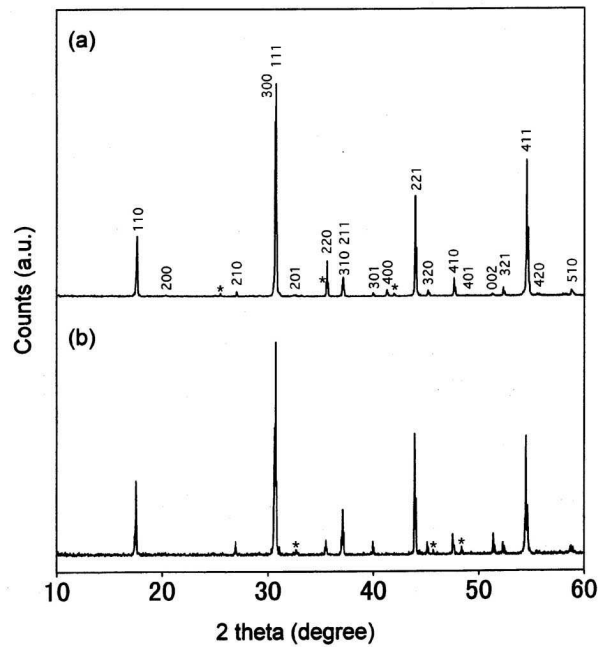}
\caption{Plots of x-ray profiles (CuK$_\alpha$) of powder samples of the 
cobalt oxide (a) annealed in the compressed oxygen-argon gas and (b) as-made
in air. The data were obtained at room temperature. Small peaks marked by 
stars have not been indexed by the hexagonal-unit-cell model.}
\label{XRD}
\end{figure*}

The powder x-ray patterns of the annealed sample and the
as-made sample are presented in Figs.\ref{XRD}a and \ref{XRD}b, respectively.
Recalled the hexagonal unit cell of the structure of the analogous copper 
oxide Sr$_5$Pb$_3$CuO$_{12}$ ({\it P}\={6}2{\it m}, $a=$ 10.1089(6) \AA~and 
$c=$ 3.5585(2) \AA) \cite{PRB01KY}, a hexagonal unit cell was, at first, tested
to qualitatively analyze the both patterns. 
Almost peaks were clearly found to be at the expected positions from the 
hexagonal symmetry and lattice parameters, as marked by each $hkl$ notation
(Fig.\ref {XRD}a), except several small peaks indicated by solid stars.
The lattice parameters were refined to $a=$ 10.11(1) {\AA} and $c=$ 3.567(1)
{\AA} by a least squares fitting for the annealed sample pattern (Fig.\ref 
{XRD}a), and $a=$ 10.12(1) {\AA} and $c=$ 3.558(1) {\AA} for the other (Fig.\ref 
{XRD}b).
The both patterns indicate quality of the samples.
Although the hexagonal model was eventually found reasonable to explain the 
almost x-ray peaks for the both cobalt oxides as well as the copper oxide, the 
star-marked minor peaks, however, remained to be uncharacterized, indicating
possible presence of either small amount of impurities otherwise somewhat 
structural modulation.
An attempt to prepare a much pure sample, which may not show the extra
peaks, by means of optimizing the heating conditions and starting 
compositions has been made, however, that is unsuccessful thus far.

\begin{figure*}
\includegraphics{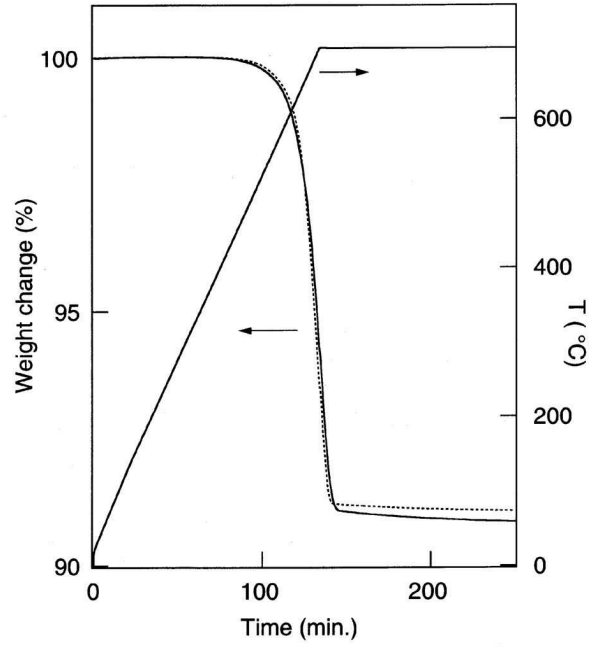}
\caption{Thermogravimetric analysis data for the powder samples 
as-made (broken curve) and annealed in the compressed oxygen-argon gas
(solid curve). The samples have been studied in a mixed gas, 3\% hydrogen in 
argon, at heating ratio of 5 $^\circ$C per minute. The weight loss at the 
major steps were 8.903 \% (solid curve) and 8.758 \% (broken curve), which 
suggesting oxygen quantity of the samples 12.32 and 12.19 moles per the formula 
unit, respectively.}
\label{TGA}
\end{figure*}

In order to measure the oxygen content of the cobalt oxide, the TGA study was made 
on the selected samples; the data are shown in Fig.\ref{TGA}, where dotted and solid 
curves indicate weight change for the as-made and annealed samples, 
respectively.
A clear weight loss was found above approximately 400 $^\circ$C in the both
measurements.
Based on the hypothesis that the observed weight loss is totally due to oxygen
reduction, the oxygen quantity was then calculated to be
12.19 and 12.32 moles per the formula unit for the as-made and annealed 
samples, respectively.
The high-oxygen pressure annealing does not appear to produce a significant 
increment of oxygen quantity in the as-made cobalt oxide.
This fact is supported by the x-ray data; rather small changes (less than
0.25\%) in lattice parameters was found after the annealing.
As the samples continue to louse those weight after the oxygen reduction 
completed (null data secured the accuracy of the measurements), probably due 
to a volatility of lead, a small amount of extra loosing should be 
superimposed on the major steps, which may lead an over estimation somewhat into 
the oxygen-quantity calculations. 

\begin{figure*}
\includegraphics{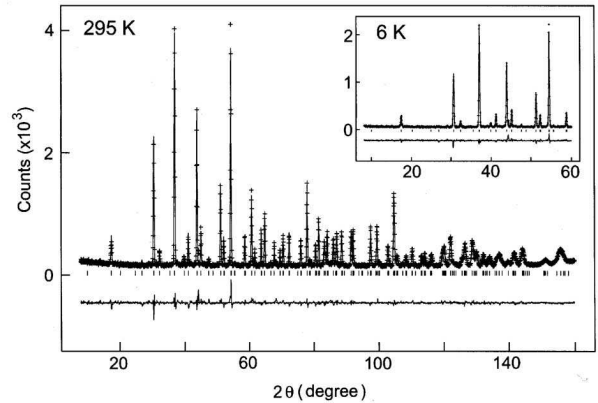}
\caption{Plots of the observed (crosses) and calculated (solid curve) neutron
diffraction profiles (295 K, $\lambda = 1.5401$ \AA) of the powder sample of
Sr$_5$Pb$_3$CoO$_{12}$, annealed in the compressed gas. The vertical bars 
indicate calculated positions for the nuclear Bragg reflections. The lower 
part shows difference between the profiles. The data at 6 K and the subsequent
analysis of those are presented in the inset.}
\label{NDP}
\end{figure*}

The annealed sample was subjected to neutron diffraction study at 6 and 295
K to obtain details of the local-chain structure and degree of
oxygen nonstoichiometry. 
The analysis of the neutron data using the Rietveld technique fundamentally
followed the way developed in the structural studies on the 
analogous copper oxide Sr$_5$Pb$_3$CuO$_{12}$ \cite
{PRB01KY,JSSC90JSK,JSSC91TGNB}.
As a result, a good achievement was successfully obtained by the Rietveld 
calculations on the neutron profiles (Fig.\ref{NDP}), suggests the average 
crystal structural model ({\it P}\={6}2{\it m}) is reliable with the cobalt 
oxide as well as the copper oxide \cite{PRB01KY,JSSC90JSK,JSSC91TGNB}.
In the main panel of Fig.\ref{NDP}, the observed and calculated profiles are
presented at the best quality (5--7 \% levels in agreement factors).
Shown below the profiles as a deference plot between those, quality of the 
refinement is indeed on a convincing level.
The structure parameters at the best quality are listed in 
Table \ref{table1}, and selected interatomic distances and bond angles were 
calculated from the parameters (Table \ref{table2}).
Lattice constants at 295 K of the hexagonal unit cell are $a =$
10.1093(2) \AA~and $c =$ 3.562 51(9) \AA, which essentially meet the x-ray parameters. 

During the refinements, temporary fits using unfixed 
isotropic-atomic-displacement parameters and occupancy factors for O(4) and 
O(5) had lead only incredible results, probably due to too low levels of those 
occupancy factors less than 0.1.
The situation was not improved even for the low temperature data.
The isotropic atomic displacement parameters at the oxygen sites O(4) and O(5)
were, therefore, fixed at a conceivable level, 2.0 \AA$^2$ (1.2 \AA$^2$ at 6 K);  
calculations thereafter appeared to improve the situation.
Due to the same reasons, the occupancy factor of O(5) was constrained to be 
equal to that of O(4).
The occupancy factors of O(2) and O(3), the normally occupied sites, were
fixed to be fully occupied in the final calculation because oxygen vacancy was
found within one standard deviation of 1.00.
The possible mixing between Sr and Pb was tested in preliminary manner, however
substantial degree of the mixing was not detected. 
The displacement parameter of Co is unusually large, probably reflecting the 
local structure disorder as in the copper oxide \cite{PRB01KY,JSSC90JSK,JSSC91TGNB}.
The occupancy factors in the partially occupied sites, O(1), O(4), and O(5),
are slightly higher than the expected values 2/3, 1/12, and 1/12, 
respectively, from the oxygen-stoichiometric composition 12 moles per the 
formula unit.
The oxygen quantity calculated from the present neutron parameters is 12.42 
moles per the formula unit.
Although the estimation from the neutron data matches that of the TGA data 
(12.32 moles) within 1 \% level, we believe the oxygen quantity may be 
slightly overestimated because the occupancy factors of O(4) and O(5) are too
low to be accurate.
As the neutron diffraction study on the copper oxide did not help to figure 
out the probable local-structure disorders and oxygen nonstoichiometry \cite
{PRB01KY,JSSC90JSK,JSSC91TGNB}, further attentions on the present analysis, 
including calculations with anisotropically thermal parameters of metals, did
not shed light on the problems, either.
The oxygen composition of the cobalt oxide might be slightly 
superstoichiometric ($\lesssim$ 12.4), but not far from the stoichiometric 
Sr$_5$Pb$_3$CoO$_{12}$.
To determine the oxygen quantity conclusively, further investigations would be
required after much high-pure sample becomes available.

\begin{figure*} 
\includegraphics{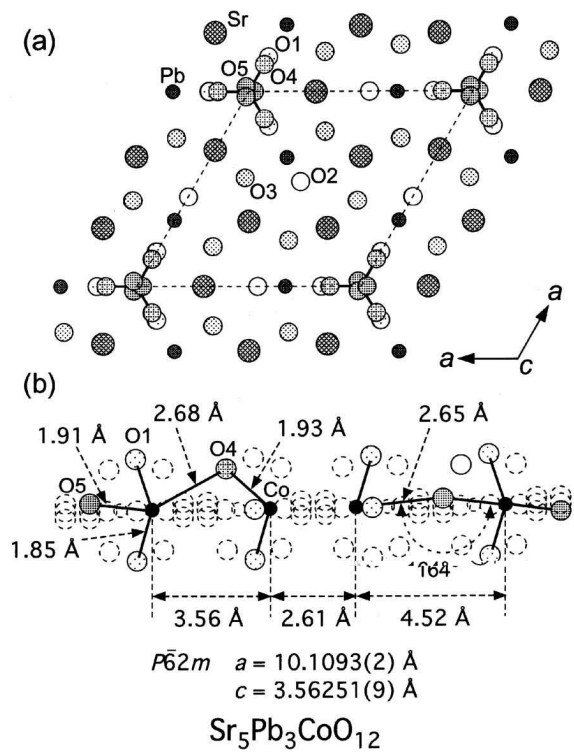}
\caption{(a) Schematic crystal structure view of Sr$_5$Pb$_3$CoO$_{12}$ 
drawn from the 295K-neutron data. The hexagonal unit cell is 
indicated by the broken lines. (b) Schematic view of the cobalt--oxygen chain
along {\it c}--axis, showing the randomly occupied oxygen positions [O(1), O(4), 
and O(5)], and a probable arrangement of cobalt and oxygen atoms within
the average structural model.}
\label{Structure}
\end{figure*}

Schematic structural views of Sr$_5$Pb$_3$CoO$_{12}$ were drawn from the 
neutron data (295 K) in Figs.\ref{Structure}a and \ref{Structure}b.
It is clear that the cobalt compound has a linear-chain structure basis, in 
which cobalt--oxygen polyhedra form chains at inter-chain distance of 10.1 
\AA~\cite{PRB01KY}.
A part of the chain along $c-$axis is presented in Fig.\ref
{Structure}b.
All of cobalt and oxygen sites are shown by the dotted circles smaller and 
larger, respectively, and a probable arrangement of those atoms is indicated
by solid and filled circles.
Intolerably short bond distances (Table \ref{table2}) between cobalt and oxygen
atoms were precluded from making the arrangement.
The irregular arrangement of cobalt atoms, as shown in Fig.\ref{Structure}b,
and probable presence of
local displacements of atoms may account for the observed unusually large 
atomic displacement parameters in the average-structure data such as 3.8 \AA$^2$
and 2.9 \AA$^2$ for Co and O(1), respectively.
The $c-$axis constant (3.56251(9) \AA) reflects the average of Co to Co 
distances in the chains; almost are close to $\sim$3.56 \AA,~but few parts at
$\sim$2.61\AA~may be somewhat involved.
Since the oxygen-coordination environment of all cobalt atoms is not unique, 
a degree of magnetic uniformity of the present chain compound might be as low
as that of the copper oxide Sr$_5$Pb$_3$CuO$_{12}$, far rather than other 
well-studied spin-chain compounds \cite {PR64JCB,PRB83JVCB,PRB87DCJ,PRB94TB}.

\begin{figure*} 
\includegraphics{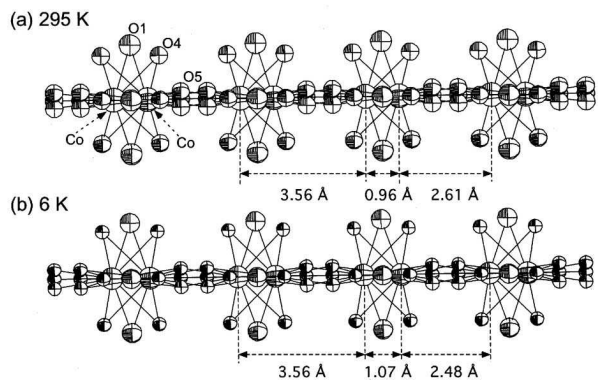}
\caption{Comparison of the chain structures at (a) 295 K and (b) 6 K of 
Sr$_5$Pb$_3$CoO$_{12}$ along {\it c}--axis. The isotropic atomic displacements
are shown. }
\label{Chain}
\end{figure*}

The low-temperature structure was investigated at 6 K (inset of Fig.\ref{NDP})
in the same manner, and the obtained structure parameters, calculated 
interatomic distances and bond angles are summarized in Tables \ref{table1} 
and \ref{table2} as well.
The hexagonal unit cell slightly shrinks (0.21 \% along $a-$axis and 0.17 \%
along $c-$axis) by the cooling, while the Co site splitting is much 
conspicuous than that at 295 K (+11.5 \% in splitting distance) as shown in 
Figs.\ref{Chain}a and \ref{Chain}b.
Any traces of magnetic ordering were not detected in the diffraction profiles. 

\begin{figure*} 
\includegraphics{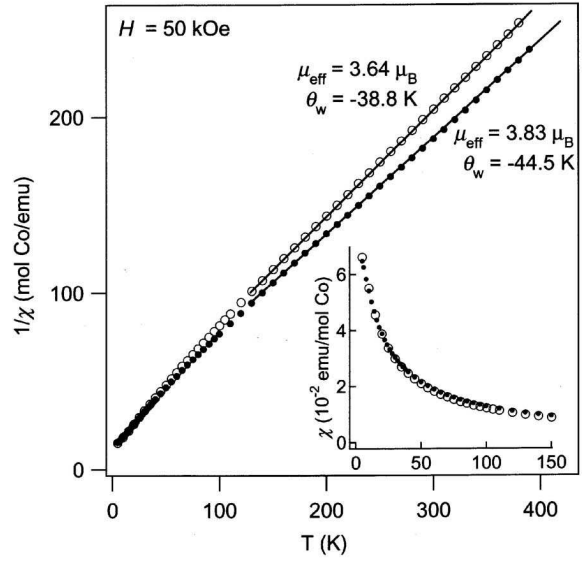}
\caption{Temperature dependence of the inverse magnetic susceptibility and the
magnetic susceptibility at 50 kOe of the cobalt oxides as-made 
(closed circles) and annealed in the compressed oxygen-argon gas (open 
circles). The solid line indicates fits to the Curie-Weiss law.}
\label{Magnetic}
\end{figure*}

\begin{figure*} 
\includegraphics{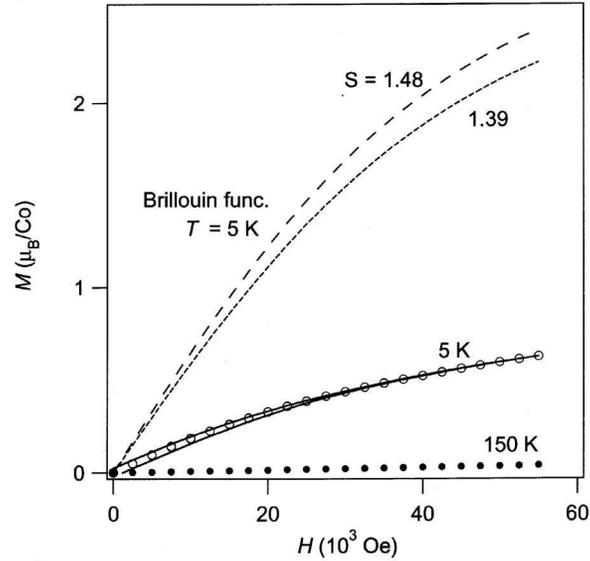}
\caption{Applied magnetic field dependence of the magnetization of the 
annealed sample of Sr$_5$Pb$_3$CoO$_{12}$ at 5 and 150 K. The solid curves are
the data at 5 K obtained prior to the annealing. Estimated spin numbers ($S 
=$ 1.48 and 1.39) by the Curie-Weiss fits to the high temperature 
data were employed to compute the broken curves for free spins at 5 K using
the Brillouin function.}
\label{MH}
\end{figure*}

The magnetic-susceptibility data of the samples annealed (open circles) and 
as-made (closed circles) are shown in Fig.\ref{Magnetic}; $T$ vs $1/\chi$ and
$T$ vs $\chi$ are plotted in the main panel and the inset, respectively.
The data of magnetic field dependence of the magnetization at 5 and 150 K are
shown in Fig.\ref{MH}.
Seen in the main panel of Fig.\ref{Magnetic}, the inverse magnetic 
susceptibility is linearly dependent on temperature above approximately 130 
K for both samples.
The CW law was, then, applied to analyze the two sets of the high-temperature
portion of the data.
The formula to fit the data by a least-squares method, as indicated by the
solid lines (main panel of Fig.\ref{Magnetic}), was
\begin{eqnarray}
\chi(T)={N\mu^2_{\rm eff} \over 3k_{\rm B}(T-\theta_{\rm W})}+\chi_{\rm 0},
\label{CW}
\end{eqnarray}
where $\theta_{\rm W}$, $\chi_0$, $\mu_{\rm eff}$, $k_{\rm B}$, and $N$
were Weiss temperature, temperature-independent term, effective magnetic 
moment, Boltzmann constant, and Avogadro's constant, respectively.
In a preliminary fit, the temperature-independent term $\chi_{\rm 0}$ was 
estimated to be on the order of 10$^{-6}$ emu/mol-Co, which was negligibly 
small, and then we decided not to employ the second term in Eq.\ref{CW} 
hereafter.
The subsequent fits yielded $\mu_{\rm eff}$ of 3.64 $\mu_{\rm B}$ per Co and
$\theta_{\rm W}$ of -38.8 K for the annealed sample, and $\mu_{\rm eff}$ of 
3.83 $\mu_{\rm B}$ per Co and $\theta_{\rm W}$ of -44.5 K for the as-made 
sample, those parameters were characteristic of antiferromagnetic interaction.
As the oxygen stoichiometry is approximate at about 3\% level (12 to 12.4 
moles per the formula unit), the formal valence of oxygen also becomes 
approximate about +2 to +2.8, when the valence of lead is +4.
Due to the rather large degree of uncertainty of the valence of cobalt at 
about 40\% level, it is unable to determine the electronic configuration of 
cobalt solely from the magnetic susceptibility data.
Further investigations, including spectroscopic studies, might help to make 
clear the issue. 

Seen in Fig.\ref{Magnetic}, a characteristic maximum expected from 1D 
antiferromagnetism in $T$ vs $\chi$ plot for either Bonner-Fisher or spin-gap
models was not observed at all, although the chains were substantially 
antiferromagnetic \cite{PR64JCB,PRB83JVCB,PRB87DCJ,PRB94TB}.
Below approximately 100 K, the inverse magnetic susceptibility gradually 
starts to apart from the CW line to go down on cooling, indicative of 
gradual growth of a ferromagnetic component in short range.
To study the low-temperature portion further, the Brillouin function was 
employed to make comparison with the magnetization data at 5 K (Fig.\ref{MH}).
The function to calculate the ideal $M_S$ vs $H$ curve at 5 K for free spins
($S$) was
\begin{eqnarray}
\lefteqn{M_S(\alpha)=NgS\mu_{\rm B}} 
\nonumber
\\
&&\times\Bigg( \frac{2S+1}{2S}\coth\frac{2S+1}{2S}\alpha- \frac
{1}{2S}\coth\frac{1}{2S}\alpha \Bigg),
\label{Brillouin}
\end{eqnarray}
where $\alpha = gS\mu_{\rm B}H/k_{\rm B}T$. 
The effective $S$ was estimated to be $\sim$1.39 for the annealed and 
$\sim$1.48 for the as-made samples from calculations using the formula $\mu_
{\rm eff} = 2\sqrt{S(S+1)}$ and the $\mu_{\rm eff}$ parameters, obtained by
the CW fits. 
The computed curves for free spins at the both $S$ vales are shown in Fig.\ref{MH}.
Due to the substantial antiferromagnetic interactions, the magnetization at 
50 kOe for the cobalt oxides is reduced to approximately one third of that for
the ideal-free-spin system.
Although the absolute Weiss temperatures are much higher than the studied 
temperature 5 K, the magnetization depends on applied field with rather 
positive curvature than linear relationship, which may indicate somewhat an 
influence of the ferromagnetic spin coupling (Fig.\ref{MH}).
These observations, including the deviation from the CW line on cooling, 
suggest that minor ferromagnetic interactions coexist with the 
antiferromagnetism, and might help to destroy the potential 
1D-antiferromagnetic characters.
Since the chain involves local-structure disorder and potentially be 
modulated, characters of all probable magnetic bonds were 
unable to be uniquely determined in this way. 
The observed $\theta_{\rm W}$, which is apparently negative, is probably due
to a balance between the major antiferromagnetic and the possible minor 
ferromagnetic interactions.
The observed ferromagnetic component, otherwise, results from minor magnetic
impurities.

In summary, we have reported the average structure and the magnetic properties
for the quasi-1D novel cobalt oxide.
Due to the probable presence of local-structure disorders such as the local 
displacement of cobalt and ligand oxygens, and the irregular arrangement of 
cobalt-oxygen polyhedra in short range, degree of magnetic coherence should 
not be as high as expected to the low dimensional magnetism.
The local structure is too complicated to be clearly observed by the neutron
diffraction which probes a positional average of the structure.
Further studies with local structural and magnetic probes, more sensitive to
the microscopic magnetic environment, could clarify the probable coupling 
between the local-structure disorders and the rather unusual magnetism.

\acknowledgments
This research was supported in part by the Multi-Core Project administrated 
by Ministry of Education, Culture, Sports, Science and Technology of Japan.

\begin{table*}
\caption{Structure parameters of Sr$_5$Pb$_3$CoO$_{12}$ at 295 K (first line)
and 6 K (second line). Space group: {\it P}\={6}2{\it m}. The lattice 
parameters are {\it a} = 10.1093(2) \AA, {\it c} = 3.562 51(9) \AA\ at 295 K,
and {\it a} = 10.0877(2) \AA, {\it c} = 3.555 87(9) \AA\ at 6 K.~~The volume
of the hexagonal unit cell is 315.31(2)  \AA$^3$~at 295 K and 313.38(2) \AA$^3$
at 6 K.~~The calculated density is 6.89 g/cm$^3$ at 295 K and 6.93 g/cm$^3$ 
at 6 K.}
\label{table1}
\begin{ruledtabular}
\begin{tabular}{llllllll}
Atom &Site &$x$         &$y$         &$z$         &$n$          &$B$ (\AA$^2$)&\\
\hline
Sr(1)&$2d$ &1/3         &2/3         &1/2         &1            &1.302(86)    &\\
     &     &1/3         &2/3         &1/2         &1            &0.617(81)    &\\
Sr(2)&$3g$ &0.700 79(25)&0           &1/2         &1            &0.829(51)    &\\
     &     &0.701 26(27)&0           &1/2         &1            &0.519(52)    &\\
Pb   &$3f$ &0.338 42(19)&0           &0           &1            &0.897(29)    &\\
     &     &0.338 27(20)&0           &0           &1            &0.545(31)    &\\
Co   &$2e$ &0           &0           &0.366(7)    &0.5          &3.80(56)     &\\
     &     &0           &0           &0.349(8)    &0.5          &3.17(53)     &\\
O(1) &$3g$ &0.1764(5)   &0           &1/2         &0.795(23)    &2.94(25)     &\\
     &     &0.1779(6)   &0           &1/2         &$= n$(295 K) &2.47(13)     &\\
O(2) &$3g$ &0.461 17(35)&0           &1/2         &1            &1.320(65)    &\\
     &     &0.4617(4)   &0           &1/2         &1            &0.765(62)    &\\
O(3) &$6j$ &0.237 57(22)&0.4418(4)   &0           &1            &0.943(41)    &\\
     &     &0.237 97(24)&0.4428(4)   &0           &1            &0.662(43)    &\\
O(4) &$6i$ &0.1419(30)  &0           &0.271(8)    &0.086(5)     &2            &\\
     &     &0.1499(31)  &0           &0.282(7)    &$= n$(295 K) &1.2          &\\
O(5) &$6i$ &0.9746(28)  &0           &0.896(5)    &$= n$[O(4)]  &2            &\\
     &     &0.9696(24)  &0           &0.896(5)    &$= n$(295 K) &1.2          &\\
\hline
 &$R_{\rm p} =$&5.12 \%&$R_{\rm wp} =$&6.66 \%&$\chi^{2} =$&1.970&\\
 &             &6.79 \%&              &8.60 \%&            &1.428&\\
\end{tabular}
\end{ruledtabular}
\end{table*}

\begin{table*}
\caption{Selected interatomic distances and angles of Sr$_5$Pb$_3$CoO$_{12}
$ at 295 K (first line) and 6 K (second line).}
\label{table2}
\begin{ruledtabular}
\begin{tabular}{lllllll}
Atoms         &            &Distances (\AA)&Atoms         &            &Distances (\AA)&\\
\hline
Sr(1)--O(2)   &$\times$3   &2.9446(5)      &Pb--O(3)      &$\times$2   &2.0857(19)     &\\
              &            &2.9376(5)      &              &            &2.0841(21)     &\\ 
Sr(1)--O(3)   &$\times$6   &2.6600(24)     &Pb--O(4)      &$\times$2   &2.209(29)      &\\
              &            &2.6483(25)     &              &            &2.149(30)      &\\ 
Sr(2)--O(1)   &$\times$2   &2.6335(21)     &              &$\times$2   &3.270(31)      &\\
              &            &2.6257(23)     &              &            &3.291(11)      &\\ 
Sr(2)--O(2)   &$\times$1   &2.422(4)       &Pb--O(5)      &$\times$4   &3.321(13)      &\\
              &            &2.417(5)       &              &            &3.291(11)      &\\
Sr(2)--O(3)   &$\times$4   &2.5571(25)     &Co--O(1)      &$\times$3   &1.847(9)       &\\
              &            &2.5574(27)     &              &            &1.873(10)      &\\
Sr(2)--O(4)   &$\times$4   &2.745(9)       &Co--O(4)      &[$\times$3  &1.474(30)]     &\\
              &            &2.722(8)       &              &[           &1.531(31)]     &\\
Sr(2)--O(5)   &$\times$2   &3.106(26)      &              &$\times$3   &1.932(32)      &\\
              &            &3.052(23)      &              &            &2.001(34)      &\\
              &$\times$2   &3.461(16)      &              &$\times$3   &2.68(4)        &\\
              &            &3.476(15)      &              &            &2.71(4)        &\\
Pb--O(1)      &$\times$2   &2.420(4)       &Co--O(5)      &[$\times$3  &1.694(33)]     &\\
              &            &2.404(4)       &              &[           &1.640(35)]     &\\
Pb--O(2)      &$\times$2   &2.1709(21)     &              &$\times$3   &1.905(33)      &\\
              &            &2.1704(22)     &              &            &1.968(35)      &\\
              &            &               &              &$\times$3   &2.645(30)      &\\
              &            &               &              &            &2.701(32)      &\\
\\
Atoms         &            &Angles ($^\circ$)&Atoms       &            &Angles ($^\circ$)&\\
\hline
O(1)--Co--O(1)&            &113.5(6)       &O(4)--Co--O(5)&            &44.5(11)       &\\
              &            &112.2(7)       &              &            &45.1(11)       &\\
O(1)--Co--O(4)&            &121.96(2)      &O(4)--Co--O(5)&            &55.3(21)       &\\
              &            &121.18(32)     &              &            &58.1(14)       &\\
O(1)--Co--O(5)&            &71.2(8)        &O(4)--Co--O(5)&            &121.0(10)      &\\
              &            &69.1(8)        &              &            &124.9(11)      &\\
O(1)--Co--O(5)&            &82.7(12)       &O(4)--Co--O(5)&            &147.5(21)      &\\
              &            &82.3(12)       &              &            &150.2(21)      &\\
O(1)--Co--O(5)&            &100.7(10)      &Co--O(5)--Co  &            &155.9(26)      &\\
              &            &101.0(9)       &              &            &149.9(23)      &\\
O(1)--Co--O(5)&            &120.4(20)      &Co--O(5)--Co  &            &163.6(18)      &\\
              &            &126.0(20)      &              &            &160.3(15)      &\\
\end{tabular}
\end{ruledtabular}
\end{table*}

\end{document}